\documentclass[aps,twocolumn,superscriptaddress,nofootinbib]{revtex4} 

\newcommand{\be}{\begin{equation}}
\newcommand{\ee}{\end{equation}}

\newcommand{\crec}{\chi_{\mbox{\scriptsize{rec}}}}

\usepackage{graphicx}
\usepackage{color}
\usepackage{dcolumn}
\usepackage{bm}
\usepackage{amsmath}
\usepackage[unicode=true,pdfusetitle,
 bookmarks=true,bookmarksnumbered=false,bookmarksopen=false,
 breaklinks=true,pdfborder={0 0 0},pdfborderstyle={},backref=false, colorlinks=true, citecolor = blue]
 {hyperref}
\usepackage{amssymb}
\usepackage{multirow} 
\usepackage{physics} 
\usepackage{enumitem}
	\setlist{noitemsep} 

\begin{document}

\title{Topology and the suppression of CMB large-angle correlations}

\author{Armando Bernui}
\email[]{bernui@on.br}
\author{Camila P. Novaes}
\email[]{camilapnovaes@gmail.com}
\affiliation{Observat\'orio Nacional, Rua General Jos\'e Cristino 77,   \\
S\~ao Crist\'ov\~ao, 20921-400 Rio de Janeiro -- RJ, Brazil} 
%
%
\author{Thiago S. Pereira}
\email[]{tspereira@uel.br}
\affiliation{Departamento de F\'{\i}sica, Universidade Estadual de Londrina, 
Rodovia Celso Garcia Cid, km 380, 86051-990, Londrina -- PR, Brazil}
\author{Glenn D. Starkman}
\email[]{glenn.starkman@case.edu}
\affiliation{CERCA/ISO/Physics Department, Case Western Reserve University, 
Cleveland, OH 44106-7079, USA}

\date{\today}

\begin{abstract}
To date, no compelling evidence has been found 
that the universe has non-trivial spatial topology. 
Meanwhile, anomalies in the observed
cosmic microwave background (CMB) temperature map, 
such as the lack of correlations at large angular separations, 
remain observationally robust (e.g. \cite{Schwarz:2015cma}).
We show that if our universe is flat and has one compact dimension
of appropriate size (a.k.a. slab or $R^2\times S^1$ topology),
this would suppress large-angle temperature correlations 
while maintaining a low-$\ell$ angular power spectrum 
consistent with observations. The optimal length appears to be $1.4$ times 
the conformal radius of the CMB's last scattering surface ($\crec$).
We construct the probability distribution function of the statistic $S_{1/2}$ 
using 1000 simulated Sachs-Wolf-only skies for each of  several values of $L_z/ \crec$,
while retaining the $3$-d Fourier-mode power-spectrum 
of the best-fit $\Lambda$CDM cosmology. For $L_z\simeq1.4\crec$
the $p$-value of four standard masked Planck maps is $p \simeq 0.15$, 
compared to $p\lesssim 0.003$ 
for the conventional topologically trivial (covering) space. 
The mean angular power spectrum $\langle C_{\ell} \rangle$
of the $L_z= 1.4\crec$ slab space
matches the observed full-sky power spectrum at $2\leq\ell\lesssim6$ 
-- 
including a substantially suppressed  quadrupole $C_2$,
a slightly suppressed octopole $C_3$, 
and unsuppressed higher multipoles.
It does not predict other low-$\ell$ CMB anomalies,
and does not take account of normally sub-dominant 
Integrated Sachs Wolfe (ISW) contributions. 
An $L_z= 1.4\crec$ slab topology 
is consistent with published limits from the Planck maps~\cite{PLA-topology-2015},
which require only $L_z\gtrsim 1.12\crec$.
It is within the $95\%$ confidence range $1.2\leq L_z/ \crec\leq 2.1$ 
inferred using the covariance-matrix of temperature fluctuations~\cite{AM}. 
However, it violates published circles-in-the-sky limits from WMAP~\cite{WMAPtopologycontraints}
and related unpublished limits from Planck \cite{Vaudrevange:private}, 
which require $L_z/ \crec\gtrsim1.9$.
We remark on the possibility to satisfy these limits,
and  ``postdict'' other large-angle anomalies, 
with closely related topologies.
\end{abstract}

\maketitle

Cosmic Microwave Background (CMB) data is arguably the best tool extant
to determine the topology of the universe. 
According to  General Relativity, 
the dynamics of spacetime
are governed by the Einstein Field Equations, which are local.
Global information like topology requires observations,
like the CMB, that probe the Universe on the largest scales.

So far, direct searches for spatial topologies 
using data from WMAP~\cite{WMAPtopologycontraints}
and Planck~\cite{PLA-topology-2013,PLA-topology-2015} 
have found no convincing evidence of compact dimensions below the 
radius of the last scattering surface (LSS) $\crec$ -- 
the very nearly spherical locus of points 
from which most CMB photons we now detect 
were emitted $\sim13.8$ billion years ago.

Meanwhile,
analyses of CMB data have revealed a set of 
anomalous statistical features at large angles 
that appear incompatible with the null hypothesis
that the Universe is statistically homogeneous and isotropic, 
with primordial fluctuations (generated by inflation)
characterizable as Fourier modes 
with amplitudes that are 
Gaussian-random statistically independent (GRSI) variables 
	of zero mean~\cite{Spergel03,Bennett11,PLA1-XXIII,PLA2-XVI,Copi04,Copi10,Eriksen04,StatIsoVioln,Akrami}. 
One of these anomalies, first observed 
	by the Cosmic Background Explorer (COBE) satellite~\cite{Hinshaw96}, 
concerns the low amplitude 
	of the temperature two-point angular-correlation function (2PACF)
on scales above $\sim60^{\circ}$,
compared with what is expected in 
the standard inflationary $\Lambda$CDM cosmology.
Subsequent analyses with both WMAP~\cite{Spergel03} 
	and Planck~\cite{PLA1-XXIII,PLA2-XVI} data 
confirmed this anomaly. 
The reported $p$-values quantifying the (un)likelihood 
that this observation is a statistical fluctuation 
are in the range $0.03\% - 0.3\%$, 
depending on the details of the analysis performed~\cite{Copi06to13EfstathiouHajian07Kim10Kim11Gruppuso14,Copi09,PLA1-XXIII,PLA2-XVI}. 
A convincing physical explanation appears to require 
not merely a change in the primordial 3-d power spectrum, 
but that the amplitudes of Fourier modes 
are not GRSI.

One way for the Universe to evince statistical anisotropies  
without altering the local geometry
is for it to have non-trivial topology.
Topology would enforce boundary conditions 
on the modes underlying primordial fluctuations -- 
prohibiting some Fourier modes
and imposing exact relations 
among the amplitudes of others. 
These boundary conditions generically 
break homogeneity and spherical symmetry, 
and the resulting primordial fluctuations 
reflect that symmetry breaking.
In the spherical coordinate basis most appropriate for
describing CMB observations,
correlations will be induced between coefficients $a_{\ell m}$
of spherical harmonics $Y_{\ell m}$, 
including between those of different $\ell$,
and consequently between different elements 
$C_\ell$ of the angular power spectrum (APS).  

Correlations between $C_\ell$s are what is required 
to realize the observed lack of angular correlations 
without suppressing the APS over
a wide range of $\ell$.  
We show here
that a single compact dimension in flat geometry 
could induce precisely the right correlations. 
We find that the appropriate size of that dimension,
$L_z\simeq 1.4\crec$,
is within the $95\%$ confidence range ${1.2\leq L_z/ \crec\leq 2.1}$
inferred from a Bayesian analysis 
of the full CMB temperature-temperature
correlation function performed using WMAP data \cite{AM}, 
and is consistent with published limits obtained using Planck data
\cite{PLA-topology-2013,PLA-topology-2015}.
However, $L<2\crec$ implies that the LSS self-intersects,
and consequently contradicts published \cite{WMAPtopologycontraints}
	limits $L_z\gtrsim1.9\crec$
from searches of WMAP full-sky data for the resulting matched
circles-in-the-sky (and similar unpublished limits using Planck maps 
\cite{Vaudrevange:private}). We discuss below how these strong indications of cosmic topology
might be reconciled with the absence of matched circles. 

We model our universe 
by the so-called \cite{AdamsShapiro} slab model ($T_1$, with topology $R^2\times S^1$), 
where one spatial direction is compact 
while the other two are infinite. 
The slab  is also a phenomenological stand-in \cite{PLA-topology-2015}
for a 3-torus ($T_3$),
where two of the spatial dimensions are large (${L_x,L_y \gg 2\crec}$),
or for any topology that admits a flat or curved homogeneous geometry 
in which at certain locations in the fundamental domain
the space appears to be compact only in one dimension.

While keeping all of these limits in mind, we nevertheless treat 
$L_z/ \crec$ as a free parameter and consider specifically
${L_z/ \crec \,=\{ 1.15, 1.4, 1.9\}}$. 
We generate ensembles of realizations of CMB skies for 
each of these three values of $L_z/ \crec$ and compare
to an ensemble of realizations of the covering space
$L_z/ \crec = \infty$.

Cosmological perturbations can be expanded 
in a complete basis of eigenfunctions, $Q_\mathbf{k}(\mathbf{x})$, 
of the Laplace operator with the appropriate boundary conditions.
In a spatially flat topologically trivial FLRW spacetime, 
these eigenfunctions can be written as
\be
Q_\mathbf{k}(\mathbf{x})= N_k
e^{i\mathbf{k}\cdot\mathbf{x}}\,,\qquad \mathbf{k}\in {R}^3\,,
\ee
where $N_k$ is some appropriate normalization.
In a slab space $Q_\mathbf{k}$ is of the same form, 
but, taking $z$ to be the compactified direction,
$k_z$ assumes only discrete values
\begin{equation}
\mathbf{k} = \bigg(k_x,k_y,2\pi\frac{n_z}{L_z} \bigg)\,,\quad
	(k_x,k_y)\in {R}^2,\;\, n_z\in {Z}\,.
\end{equation}

As a scalar perturbation, the gravitational potential responsible for the dominant
Sachs-Wolfe (SW) contribution to the temperature anisotropies can be expanded as
\be
\Phi(\mathbf{x})=
	\sum_\mathbf{k} \Phi(\mathbf{k})Q_\mathbf{k}(\mathbf{x})\,,
\ee
where the sum is over all allowed values of $\mathbf{k}$. The SW contribution is then written as
\be
\label{DeltaTSWPhix}
\Delta T(\hat{\mathbf{x}})_{SW} 
= \frac{1}{3}\Phi(\mathbf{x}=\crec  \hat{\mathbf{x}}) \,, 
\ee
with $\crec$ being the radius of the CMB sphere in comoving coordinates. Equation~\eqref{DeltaTSWPhix} 
neglects the transfer function, which, is nearly constant across the $\ell$-range of interest.

Using a Harrison-Zel'dovich approximation to the inflationary power spectrum $P(k)$,
we can write ${\Phi(\mathbf{k}) = \sqrt{P(k)}\phi(\mathbf{k})}$, with $\phi(\mathbf{k})$ a Gaussian 
random variable of unit variance. Thus
\begin{equation} \label{eq:phi1}
	\Delta T(\hat{\mathbf{x}})_{SW} 
		= \frac{1}{3}\sum_{\{k\}}\sqrt{P(k)}
			\sum_{\left\{\hat{\mathbf{k}}\right\} _{k}}
				e^{i k \crec \cos\theta_{\mathbf{k},\mathbf{x}}}
				\phi(\mathbf{k}) \,,
\end{equation}
where $\cos\theta_{\mathbf{k},\mathbf{x}} = \hat{\mathbf{k}}\cdot\hat{\mathbf{x}}$, and $\phi(\mathbf{k})=\phi^{*}(-\mathbf{k})$ 
enforces $\Phi$'s reality.

For a fixed map resolution, the relation between the multipole $\ell$ 
and the wave number $\mathbf{k}$ for  SW perturbations is $\ell \sim |\mathbf{k}| \crec$. 
For large-angle effects, $\ell\leq\ell_{\mbox{\footnotesize max}} = 20$ suffices.
This  is less than the $\ell_{\mbox{\footnotesize max}}=40$ 
adopted by the Planck team  in their topology-likelihood 
analyses~\cite{PLA-topology-2013,PLA-topology-2015},
but is sufficient for our qualitative considerations.

We explore the slab topology with three representative values of $L_z/\crec$, namely, model [a]
with $L_z/\crec=1.15$ -- just above the published Planck lower limit; model [b]
with $L_z/\crec=1.4$; and [c] with $L_z/\crec=1.9$ -- the published circles-in-the-sky 
WMAP lower limit. According to Planck, $L_i\geq 2.2\crec$ is indistinguishable from $L_i\to\infty$,
so we take model [d] with $L_z/\crec=4$ to represent the statistically isotropic covering space.

For each of the four values of $L_z/\crec$,
we produce $10^3$ simulated full-sky SW maps 
at {\tt HEALPix} resolution  $N_{\mbox{\footnotesize side}} = 16$ 
and $2\leq\ell\leq\ell_{\mbox{\footnotesize max}} = 20$,
i.e., we use \eqref{eq:phi1} to build pixel-space maps
then filter out $\ell<2$ and $\ell>\ell_{\mbox{\footnotesize max}}$.
In building these simulated maps, we should choose
\be 
\label{nmax}
n_i \ll \frac{\ell_{\mbox{\footnotesize max}}L_i}{2\pi\crec}
\,,
\ee
so we limit $n_i$ to the range $n_i\in(-25,25)$ in all cases.

For comparison with observations, 
we have analyzed the four foreground-cleaned CMB maps 
released by the Planck Collaboration 
in their second data release~\cite{Adam:2015tpy}, namely, the {\tt SMICA}, 
{\tt NILC}, {\tt SEVEM}, and Commander maps. These are high-resolution maps, 
with Healpix \cite{Gorski} resolution 
	$N_{\mbox{\scriptsize side}}$ = 2048.
We extracted the multipoles $2 \leq \ell \leq 20$, 
and rebuilt these maps using $N_{\mbox{\footnotesize side}}$ = 16.

The Planck team also released a variety of masks \cite{Adam:2015tpy}
associated with each of the four full-sky maps.
The {\tt UT78} mask is the union of the confidence masks 
associated with each map; 
it includes a fraction $f_{\mbox{\footnotesize sky}} = 0.78$
of the sky.
This mask was also downgraded 
	to ${N_{\mbox{\footnotesize side}}=16}$,
masking a pixel in the downgraded map 
if at least 50\% of the underlying high-resolution pixels are masked.
The resulting UT80 mask,
used in all our analyses,
has  $f_{\mbox{\footnotesize sky}} = 0.80$. 

Because the slab topology breaks statistical isotropy,
the orientation of the slab with respect to the Galaxy 
(i.e. mask) matters.
For $L_z/\crec=1.15$, $1.4$, and $1.9$
we therefore consider four illustrative cases  
with the slab's normal vector pointing in the directions 
(in galactic coordinates):
$[\textrm{i}]\;(l,b)  =(0^{\circ},90^{\circ})$
the north galactic pole (NGP);
$[\textrm{ii}]\;(l,b)  =(240^{\circ},-20^{\circ})$;
$[\textrm{iii}]\;(l,b)  =(270^{\circ},0^{\circ})$;
$[\textrm{iv}]\;(l,b)  =(240^{\circ},70^{\circ})$.

\begin{figure}[h]  
	\mbox{
	\hspace{-0.2cm}
	\includegraphics[width=1\linewidth]
	{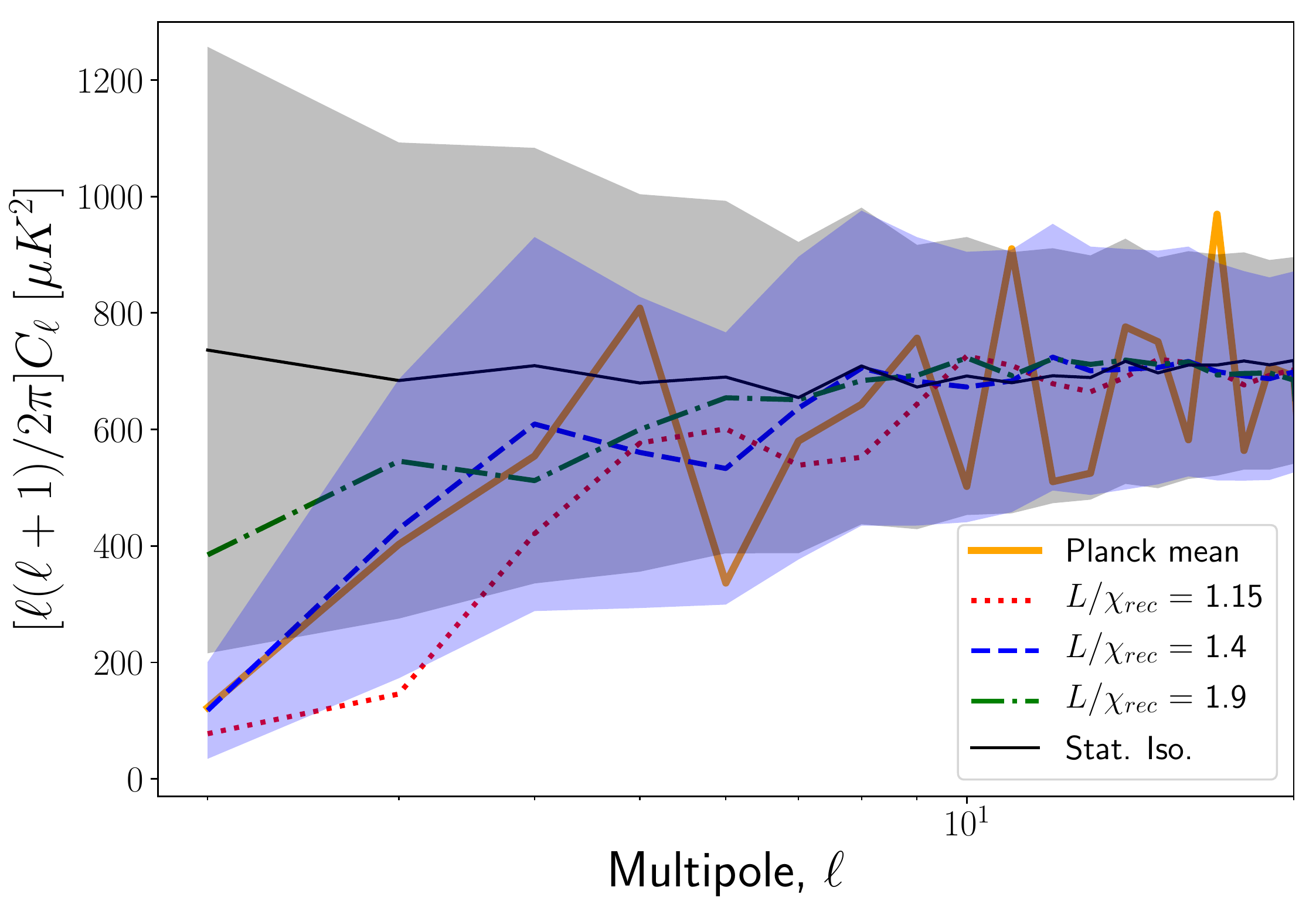}}
	\vspace{-0.7cm}
	\caption{
	Mean angular power spectrum of the data sets 
	in the slab topology model with  
	${L_z/ \crec\,=\{1.15, 1.4, 1.9\}}$ (with orientation [i]), 
	and in the statistically isotropic (SI) $\Lambda$CDM model.
	The ``Planck mean'' curve is the arithmetic mean 
	of the spectra from the four foreground-cleaned Planck maps. 
	The shaded areas correspond to ``one-sigma'' cosmic-variance bands
	around the SI (grey) and $L_z/ \crec=1.4$ (blue) models.}
\label{Clspectrum}
\end{figure}

Following \cite{Aurich08a}, 
we renormalized each set of $10^3$ slab simulations 
by a factor that matched their
mean APS in the range $15 \le \ell \le 20$ 
to that from the Planck maps. 
The renormalized APS of the 
three simulated slab-topology data sets, for direction [i],
and the statistically isotropic data set
are plotted in figure~\ref{Clspectrum},
along with the Planck APS for comparison. 
The grey and blue bands are 
the cosmic-variance bands for, respectively, the 
statistically isotropic model and  $L_z/\crec=1.4$. 
(Note, however, that the $C_\ell$ 
are not statistically independent in the slab topology.)

We wish to quantify the angular correlations 
from our  simulations, 
and compare them with one another and observations. 
The standard estimator of angular correlations 
in a map is the 
two-point angular correlation function (2PACF)~\cite{Hinshaw96,AB99,Copi10}
\be\label{eqC}
	C(\theta)\equiv
		\frac{1}{N_p}\sum_{i,j}^{Np}
			\left.\Delta T(\hat{\mathbf{x}}_i)
				  \Delta T(\hat{\mathbf{x}}_j)
				  \right|_{
				  	\hat{\mathbf{x}}_i\cdot
				  	\hat{\mathbf{x}}_j
				  	=\cos\theta
				  	} \,,
\ee
the sum running over all pairs of pixels 
separated by  
	${\theta=\arccos(\hat{\mathbf{x}}_i\cdot\hat{\mathbf{x}}_j)}$. 
Because the monopole and dipole of the CMB are large
compared to higher multipoles, 
and predominantly of different origins, 
we redefine  ${C(\theta)\to 
C(\theta)-\frac{1}{4\pi}\left(C_0 + 3 C_1 \cos\theta\right)}$.
As is conventional, if confusing, 
we refer below to this modified 2PACF as the
2PACF, and write simply $C(\theta)$.

\begin{figure}
\hspace*{-0.2cm}
\includegraphics[width=1\linewidth]
{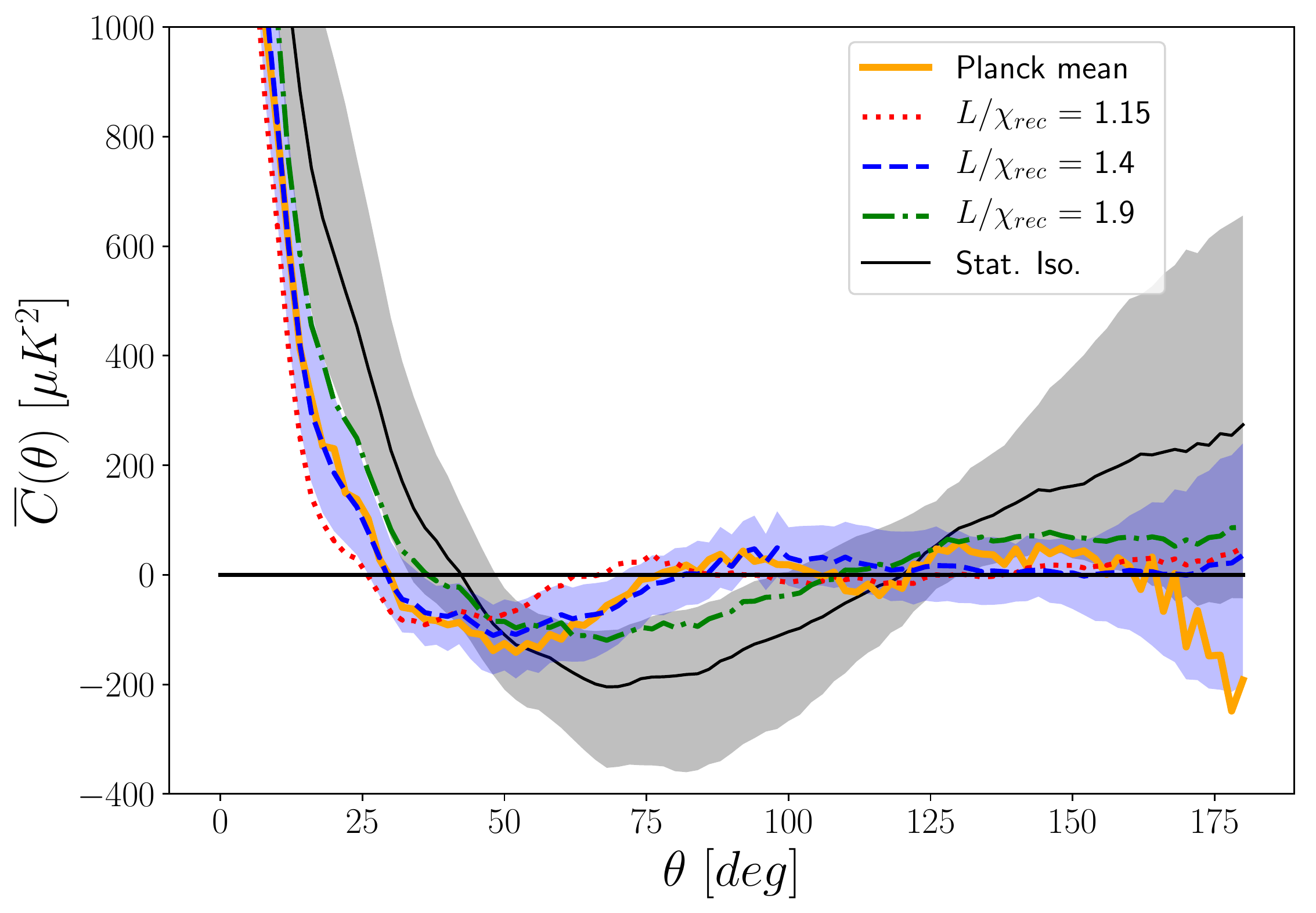}
\caption{ Mean 2PACF 
	$\overline{C}(\theta)$
from simulations of a slab topology with 
	${L_z/ \crec\,=\{1.15, 1.4, 1.9\}}$ (with orientation [i]),
and from the statistically isotropic $\Lambda$CDM simulations. 
``One-sigma'' cosmic-variance regions 
for $L_z/  \crec = 1.4$ (blue) and SI (grey) are shown. 
}
\label{Ctheta}
\end{figure}

We calculate  $C(\theta)$ for all the simulated maps
with the {\tt UT80} mask, 
obtaining thirteen ensembles each of $10^3$ functions $\{ C(\theta)\}$,
corresponding to the four values of $L_z/\crec$ 
and (for $L_z/\crec=1.15,1.4,1.9$) the four orientations.
For each of these, we compute 
the corresponding mean cut-sky 2PACF, 
	$\overline{C}(\theta)$.
We plot them in figure~\ref{Ctheta} for orientation $[i]$. (The other orientations prove nearly indistinguishable.)
We also plot the mean 2PACF of the four Planck maps. To give the reader a sense of the uncertainties, 
we include the ``one-sigma'' error bands for the statistically isotropic case (grey) and for the $L_z/\crec=1.4$ 
slab (blue); however, we caution that $C(\theta)$ for different values of $\theta$ are correlated. 

We see in figure~\ref{Ctheta} that the observed 2PACF at large angles ($\theta\gtrsim60^\circ$)
is much closer to zero than one would expect from the statistically isotropic case. 
This is the well-known ``large-angle'' angular-correlation anomaly.

Figures~\ref{Clspectrum} and~\ref{Ctheta} clearly show that the CMB angular correlations 
are sensitive to the size of the slab in the way one might expect:  
smaller slabs yield lower values of $C_\ell$ at very low-$\ell$
and result in suppressed $C(\theta)$ at large $\theta$. More specifically, 
large-angle correlations can exhibit noticeable suppression over a broad range of angular 
separations $\theta$ without strong suppressing the APS over a broad range of $\ell$.  Much like the data.

We call the reader's particular attention to the blue-dashed curves 
for $L_z/\crec=1.4$ in figures~\ref{Clspectrum} and~\ref{Ctheta}
and their similarity to the curves for the Planck mean (solid orange),
and contrasted to the statistically isotropic simulations 
(solid black).

The very-low-$\ell$ APS for $L_z/ \crec=1.4$
exhibits the same pattern as does the Planck mean --
marked quadrupole ($C_2$) suppression (relative to the SI), 
mild octopole ($C_3$) suppression,
and no noticeable suppression of $C_\ell$ for higher $\ell$.
The $L_z/ \crec=1.4$ simulations do not reproduce the observed pattern
of  odd-$\ell$ $C_\ell$ being larger than even-$\ell$ ones
known as the parity anomaly.
This is unsurprising since the slab topology has a preferred axis
($\pm\hat{z}$ -- the normal to the slab faces) 
but no preferred direction.
See below for further discussions.

The 2PACF for $L_z/ \crec=1.4$ also
exhibits remarkably close to the same pattern as does the Planck mean --
the same zero crossing at $\theta\simeq30^\circ$, 
the same approximately zero value above $\theta\simeq75^\circ$.
Qualitatively, it lacks only the anti-correlation 
	for $\theta\gtrsim160^\circ$, 
which could be connected to residual Galaxy contamination
\cite{Copi09}.

To characterize how (im)probable is the observed lack of large-angle CMB temperature correlations
in the context of any particular model, we use the well-known statistic
\begin{eqnarray}\label{S1/2}
	S_{1/2} \, \equiv \, 
	\int_{-1}^{1/2} \, \left[C(\theta)\right]^2 \, 
		{\rm d}(\cos \theta) \,. 
\end{eqnarray}  
$S_{1/2}$ was first proposed by the WMAP team~\cite{Spergel03} and  
has since been used extensively (see~\cite{Copi10} for a review). 

We compute  $S_{1/2}$ for each of the simulated maps 
and the four foreground-cleaned Planck maps~\cite{PLA2-XVI}
and construct a probability density function (PDF) 
for each value of $L_z/\crec$ and each slab orientation.
The $p$-values of the $S_{1/2}$ statistic for each topology with slab orientation [i], 
relative to each of the four Planck maps, is listed in Table~\ref{table1}. 
In Table~\ref{table2}
we show the $p$-values of the four topologies in comparison with $S_{1/2}^{\tiny{\mbox{\tt SMICA}}}$, for the four 
chosen directions ([i]-[iv]).

For the simply connected (statistically isotropic) case [d],
these $p$-values are consistent with those found in~\cite{PLA2-XVI}
and quite small, $p\simeq0.3\%$.
Smaller $L_z$ increases that $p$.

The smallest slab ($L_z=1.15\crec$) 
produces quite high $p$-values ($p\simeq40\%$)
but appears, at least by eye to over-suppress
large-angle $C(\theta)$ and low-$\ell$ $C_\ell$.
The largest slab ($L_z=1.9\crec$) 
produces significant gains in the $S_{1/2}$
$p$-value compared to the covering space 
($p\simeq3\%$),
but seems to under-suppress $C(\theta)$ and low-$\ell$ $C_\ell$.
$L_z=1.4\crec$ appears to be the Goldilocks value -- 
just the right  suppression of both $C(\theta)$ and $C_\ell$,
and a credible $p\simeq15\%$ for the observed $S_{1/2}$.
We caution the reader that a simple chi-square test
is inconclusive in deciding among the models for $C(\theta)$.

\begin{table}
	\begin{tabular}{| c | c | c | c | c | c |} 
	\hline
	\multirow{2}{*}{CMB maps} &  \multirow{2}{*}{$\log[S_{1/2}]$} & \,$p$ (\%) & \,$p$ (\%) & \,$p$ (\%)  & \,$p$ (\%) \\
	\,\,&\,\, \,\,&\,\, $[a]$ \,\,&\,\, $[b]$ \,\,&\,\, $[c]$ \,\,&\,\, $[d]$  \\
	\hline \hline
	{\tt SMICA}                & 3.39  &  40.0  &  15.2  & 3.3  & 0.3 \\  
	{\tt SEVEM}              & 3.40  &  41.9  &  15.7  & 3.6  & 0.3 \\  
	{\tt NILC}                   & 3.42  &  43.9  &  17.0  & 3.8  & 0.3 \\  
	{\tt Commander}        & 3.43  &  45.1  &  17.5  & 4.1  & 0.4 \\ 
	\hline 
	\end{tabular} 
	\caption{$p$-values for the PDF of the $S_{1/2}$ values 
	obtained comparing the four foreground-cleaned 
	Planck maps~\cite{Adam:2015tpy} to ensembles of $10^3$ simulations for each of
	the ${L_z/\crec=(1.15,1.4,1.9,\infty)}$ (respectively models [a], [b], [c] and [d].)
} 
\label{table1}
\end{table}


\begin{table}
	\begin{tabular}{| c | c | c | c | c |} 
	\hline
	\multirow{2}{*}{Orientation $\backslash$ data sets} & \,$p$ (\%) & \,$p$ (\%) & \,$p$ (\%)  & \,$p$ (\%) \\
	\,\,&\,\, $[a]$ \,\,&\,\, $[b]$ \,\,&\,\, $[c]$ \,\,&\,\, $[d]$  \\
	\hline \hline
	$[\mbox{i}]$            &  40.4  &  15.2  & 3.3  & 0.3  \\  
	$[\mbox{ii}]$           &  43.2  &  14.3  & 4.3  & 0.3  \\  
	$[\mbox{iii}]$          &  39.1  &  12.2  & 3.5  & 0.3  \\  
	$[\mbox{iv}]$          &  36.4  &  10.8  & 3.1  & 0.3  \\ 
	\hline
	\end{tabular}
	 \caption{$p$-values for $S_{1/2}$ 
	of the {\tt SMICA} map  among simulated ensembles
	for  $L_z/\crec=(1.15,1.4,1.9,\infty)$ (i.e., models [a], [b], [c] and [d]);
	and for the four orientations specified in the text. 
	}
\label{table2}
\end{table}

A Bayesian analysis carried out by the Planck team 
with their 2015 data found that, for the slab topology,
$L>1.12\crec$ 
	at the $99\%$ confidence level~\cite{PLA-topology-2015}. 
Despite including 
a back-to-back matched-circles search~\cite{CSScirclesig}
applicable to the slab topology,
this Planck limit is unexpectedly weaker 
than the equivalent WMAP limit of $L_z\gtrsim1.9\crec$~\cite{
	WMAPtopologycontraints}.
However, the Planck circle search, 
used a sky with approximately $25\%$ of the pixels masked -- 
mostly near the Galactic plane.
Such masking is typically dictated
by the presence of Galactic foregrounds. 
The matched circles (or other correlations) of a slab topology 
are easily hidden in the masked region.
In contrast, the WMAP search recognized that the circle search 
relies predominantly on the Sachs-Wolfe effect at the LSS,
which is dominated by contributions in the first Doppler peak of 
the CMB at $\ell\simeq200$.  
This small-angular-scale signal is believed to 
be accurately represented in full-sky foreground-cleaned maps,
so~\cite{WMAPtopologycontraints}
 used the full-sky 
Independent Linear Combinations (ILC) map, 
originally developed precisely for this search.
The same search was repeated for the Planck 2013 maps,
and a marginally more stringent 
unpublished limit was obtained~\cite{Vaudrevange:private}.
A $L_z/ \crec=1.4$ slab topology is thus firmly ruled out.

Meanwhile, the slab topology also does not explain 
other large-angle anomalies --
the parity anomaly~\cite{Kim:2010gf},
the dipole asymmetry~\cite{Eriksen04} 
(a.k.a. low northern variance~\cite{Akrami,ODwyer:2016xov}),
quadrupole-octopole planarity  
	and alignment~\cite{deOliveira-Costa:2003utu}.

{\it Conclusions:} Large-angle CMB anomalies discovered in COBE or WMAP data
persist after Planck. 
No satisfactory explanation has yet been proposed. 
We have explored the lack of large-angle correlations
as a possible manifestation 
of a compact direction in the Universe,
as modeled by the slab topology $R^2\times S^1$.
We considered ensembles of simulated CMB maps 
consistent with a single compact dimension of size
$L_z/ \crec=1.15, 1.4, 1.9$ relative to the LSS. 
We compared their ability to 
reproduce Planck observations with simulations of 
the flat covering space.  

In a slab topology, the Sachs-Wolfe temperature fluctuations 
were able to reproduce the observed suppression
of large-angle correlations 
(small $C(\theta\gtrsim60^\circ)$ leading to small $S_{1/2}$) 
simultaneously with the observed low-$\ell$ angular power spectrum, 
with its heavily suppressed quadrupole, 
mildly suppressed octopole,
and unsuppressed higher multipoles 
(relative to the covering space).
A slab space with $L_z/\crec=1.4$ best reproduced the qualitative
features of $C(\theta)$ and $C_\ell$ seen in Planck maps,
and increased the $p$-value of $S_{1/2}$ from
the covering space's $\sim0.3\%$ to  $\sim15\%$.
The ISW will not spoil low $S_{1/2}$~\cite{Copi:2016hhq}. 

Yet, a $L_z= 1.4\crec$ compact dimension is in conflict with limits 
from WMAP and Planck.  
Also, this topology would not explain
other statistically significant large-angle anomalies.
This suggests that if topology is indeed the explanation 
of the lack of large angle correlations, 
the vanilla slab may not be the full story. 

There are simple variations on the slab topology in flat geometry
that might build on the slab's success and mitigate its failures.  
Wider exploration would allow for rotations 
(about the slab normal) or transverse rotations, 
either of which could relieve the matched-circles constraint;
they might also incorporate identifications in the dimensions transverse
to the slab normal to help explain other anomalies.
There are also topologies of curved space that 
for large curvature radius
can look  like a slab for a well-placed observer \cite{Weeks:2002jz}. 
These may be intriguing, 
given the preference for small positive curvature 
($\Omega_k\simeq 0.04\pm0.03$) in high-redshift Planck data
(TT+TE+EE+lowE) \cite{Aghanim:2018eyx}.

The agreement between the slab topology and observations at the
largest angles and lowest $\ell$ is highly suggestive.
We may be seeing the first evidence that the Universe
is not infinite, or at least not infinite in all directions.
A more intensive search for cosmic topology is merited.

\noindent
We acknowledge a PVE project from CAPES 
	(Science without Borders program 88881.064966/2014-01). 
CPN and AB acknowledge fellowships from the Brazilian Agencies FAPERJ and CNPq, 
respectively. 
TSP thanks the Conselho Nacional de Desenvolvimento Cient\'ifico 
	e Tecnol\'ogico (grant \#311732/2015-1) 
	and Funda\c c\~ao Arauc\'aria (PBA 2016) 
	for their  support. 
GDS is supported by Department of Energy grant DE-SC0009946 to CWRU,
and thanks the Observat\'orio Nacional for its hospitality.
This paper made use of observations obtained 
	with Planck, 
	an ESA science mission 
	funded by ESA Member States, NASA, and Canada. 
Some of the results in this paper have been derived using 
	the {\tt HEALPix}/healpy package~\cite{Gorski}. 



\end{document}